\documentstyle[preprint,aps]{revtex}
\input epsf.tex
\def\DESepsf(#1 width #2){\epsfxsize=#2 \epsfbox{#1}}
\begin{document}

\preprint{\vbox{\hbox{OITS-608}\hbox{hep-ph/9608231}}}
\draft
\title {SUSY GUTs contributions and model independent extractions of CP phases}
\author{\bf N. G. Deshpande, B. Dutta, and Sechul Oh }\address{ Institute of
Theoretical Science, University of Oregon, Eugene, OR 97403}
\date{August, 1996)\\(To appear in Physical Review Letter}
\maketitle
\begin{abstract}We consider the origin of new phases in supersymmetric grand
unification model, and show how significant new contributions arise from the
gluino mediated diagram. We then present a more general model independent
analysis of various modes of B-decays suggested previously for measurement of
the CKM phases and point out what they really measure. It is  in principle
possible to separate out all the phases.
\end{abstract}

\pacs{PACS numbers: 13.20.He 12.15.Hh 12.60.Jv 12.10.Dm}

\newpage  We consider the origin of new CP violating phases from physics beyond
the standard model (SM) and their effect on various measurements of CKM phases
$\alpha$, $\beta$ and $\gamma$ proposed hitherto \cite{a,bb,d,e}.  Among new
sources of CP violation are multi-Higgs models 
\cite{[TL]}, the left-right model \cite{[LR]} and supersymmetry.   In this note
we focus on supersymmetry, which is very attractive from a grand unification
viewpoint and provides many new sources of CP violation. One obvious source is
the complex soft terms. Even when these are taken to be real, unification of
right handed fields, like the left handed ones, can lead to a new source of CP
violation. For example, a group like SO(10) \cite{[dh],[BHS],[AS]} or models
with intermediate gauge groups \cite{[dkd],[ddk]} like
$SU(2)_L\times SU(2)_R\times SU(4)_c$,
$SU(2)_L\times SU(2)_R\times SU(3)_c\times U(1)_{B-L}$ have these extra phases.
Supersymmetric contributions with new phases can be as large as the SM in the
$B-{\bar B}$ mixing and loop processes that lead to $b\rightarrow s{\bar q} q$. 

In this paper we make the first complete calculation of the gluino contribution
to $\Delta m_{B}$ in a SUSY grand unified S0(10) theory. This calculation can
easily be extended to the models with the intermediate gauge symmetry breaking
scale considered in references \cite{[dkd],[ddk]}. This calculation has been
done previously by assuming same masses for the SUSY particles only for the low
$\tan\beta$ scenario \cite{[BHS]}.  We consider two scenarios: (i) where the
Yukawa couplings are unified (i.e. large
$\tan\beta$ scenario) and (ii) a low $\tan\beta $ scenario. We show how the new
contributions are large and can affect the interpretation of measurement of CKM
phases. We then discuss the specific B-decay modes needed to extract the CKM
phases even in the presence of new physics.  This discussion actually uses model
independent analysis that is valid in almost any kind of departure from the
SM.   
 
Since the soft SUSY breaking terms are gravity induced, we shall assume them to
be universal   at the scale $2.4\cdot 10^{18} GeV(M_x)$ which is the reduced
Plank scale. For simplicity we also assume the soft terms to be real. It has
been shown that a grand unified model based on SO(10), which we will use in this
paper, gives rise to flavor violating processes in both quark and lepton
sectors. Consequently lepton flavor violating processes like
$\mu\rightarrow e\gamma$ put bounds on the parameter space along with
$b\rightarrow s\gamma$ \cite{[us]}. For models with the   intermediate gauge
symmetry breaking scales, the soft terms can be universal even at the GUT scale
and still give rise to these effects. The superpotential for the Yukawa sector
at the weak scale  for the SO(10) grand unification or for the grand unifying
model with an intermediate scale  can be written as \cite{[dh],[BHS],[AS],[ddk]}:
\begin{eqnarray} W&=&Q{\bf \bar\lambda_u}{\bf U^c}H_2 +Q{\bf V^*}{\bf
\bar\lambda_d}{\bf S}^2{\bf V^{\dag}}D^cH_1+ E^c{\bf V_G^*}{\bf
\bar\lambda_L}{\bf S}^2{\bf V_G^{\dag}}LH_1\, ,
 \end{eqnarray} where {\bf V} is the CKM matrix, {\bf V$_G$} is the CKM matrix
at the GUT scale (for intermediate gauge symmetry breaking models G is replaced
by I to denote the intermediate scale)  and
${\bf S}$ is the diagonal phase matrix with two independent phases. The phases
in the right handed mixing matrix for the down type quarks and the down type
squarks can give rise to new phases in
$\Delta m_{B}$ and $\Delta m_{K}$  through the gluino contribution.

The  existing calculation \cite{[AM]} for $\Delta m_{B}$ using GUT model usually
assumes that the soft terms are universal at the GUT scale($\sim 10^{16}$ GeV).
Under that assumption it is found that charged Higgs has the dominant
contribution. But with the universal boundary condition taken at the Planck or
string scale there can be a large contribution from the gluino mediated diagram
due to the fact that the fields that belong to the third generation have
different masses compared to the other generation at the GUT scale due to the
effect of the large top Yukawa coupling which gives rise to the non-trivial CKM
like mixing matrix in the right handed sector. We first consider large
$\tan\beta$ solution. In order to have a realistic fermion spectrum and the
mixing parameters in the large
$\tan\beta$ case, we use a maximally predictive texture developed in the
reference \cite{[SOTM]}. We will look at a scenario where $\lambda_t(M_G)=1$ and
$\tan\beta=57.15$, which gives $m_t$=182 GeV and
$m_b$=4.43 GeV.  For the small $\tan\beta $ scenario we have used
$\lambda_t(M_G)=1.25$ and
$\tan\beta=2$. Above the GUT scale we use one loop RGEs for the soft terms and
the Yukawa couplings\cite{[BHS]}. Below the GUT scale we will use the one loop
RGEs in matrix form in the
$3\times 3$ generation space for the Yukawa couplings and soft SUSY breaking
parameters as found in Ref.
\cite{[AM],[SUSY]} rather than just running the eigenvalues of these matrices as
is often done. Although doing this does not provide any new information when
$\tan\beta $ is small, when $\tan\beta$ is large it allows one to know the
relative rotation of squarks to quarks and sleptons to leptons.
 In the large
$\tan\beta $ scenario to  make sure that the electroweak symmetry is broken
radiatively, we need a non zero value of the D-term usually referred as $m_D^2$,
which gets introduced at the GUT scale due to the reduction of rank in the
SO(10) \cite{[FH]}. This D-term is also bounded from above and below  by
requiring the pseudo scalar mass to be positive along with the squark and
slepton masses\cite{[us]}.

We calculate $\Delta m_{B}$ using gluino contribution and compare it with the SM
result. We have done the calculation in SO(10), though this calculation can
easily be generalized to the models with the intermediate scale and other grand
unifying models. We use the expression for $\Delta m_{K}$  given in the
reference \cite{[GGRZ]} (modified for the purpose of $B^0-{\bar B^0}$ mixing),
because these expressions use the squark mass eigenstate basis derived from the
full
$6\times 6$ mass matrices, which automatically incorporates mixing between
``right-handed" down squarks and right-handed down quarks as is inevitable with
either large
$\tan\beta$ or SO(10) grand unification. We plot the ratio $\Delta
m_{{B}_{gluino}}$/$\Delta m_{{B}_{SM}}$ as a function of $\mu $ for different
values of the gaugino masses ($m_{1/2}$) in Figure 1 for the large $\tan\beta $
case, where m$_0$ is 1 TeV for the entire plot. The gluino mass is related to
the gaugino mass by the relation
$m_{\tilde g}={{\alpha_s}\over {\alpha_G}} m_{1/2}$. We take
$\alpha_s(M_z)=0.121$ and $\alpha_G=1/23.9$ and the scale for grand unification
to be $M_G=2\cdot 10^{16}$ GeV. Also in this scenario we have three variables:
$m_0$ (the universal scalar mass),
$m_{1/2}$ (the universal gaugino mass) and the $m_D^2$ (throughout our analysis
we will assume the trilinear soft SUSY breaking scalar coupling
$A^0=0$ at the Planck scale). The upper and the lower end of each curve
correspond to the upper and the lower limit of the D-term respectively. As
mentioned in the reference\cite{[us]}, the parameter space with
$m_0$ less than 1 TeV as well as $\mu >0$ is restricted by the flavor changing
neutral currents. In Figure 2 (small $\tan\beta$ case) we plot r($\equiv\Delta
m_{{B}_{gluino}}$/$\Delta m_{{B}_{SM}}$) as a function of the gaugino mass
($m_{1/2}$) for different values of the scalar masses $m_0$, where $\tan\beta$
is assumed to be 2 and $\mu<0$. In the plot   we have used the absolute value of
$\Delta m_{B_{d}}$. We restrict ourselves to the parameter space allowed by the
other flavor changing decays. We also make sure that
$\mu$ is less than 800 GeV to avoid fine tuning. In both figures the SUSY
contribution can be comparable to the SM. As a matter of fact in this parameter
space the gluino contribution to the $b\rightarrow s\gamma $ is also large
\cite{[uus],[us]}. From the graph one can see that for the scalar mass (or
the right handed slepton mass) m$_0$=1 TeV and for the gaugino mass $m_{1\over 2}
\geq 140$ GeV (or the gluino mass $\geq 405$ GeV), the SUSY contribution is small
(less than 20 $\%$) compared to the SM  in the large
$\tan\beta$ scenario. In the low $\tan\beta $ scenario, for the gaugino mass
$m_{1\over 2} \geq 200$ GeV (or the gluino mass $\geq 578$ GeV) and  for the
scalar mass (or the right handed slepton mass) m$_0\geq 200$ GeV, the SUSY
contribution becomes small (less than 20 $\%$) compared to the SM.  For a
complete SUSY calculation, there could be contributions from charged Higgs,
chargino and neutralino. The charged Higgs contribution does not change
significantly with the new boundary condition and has been found to be
comparable to or even greater than the SM contribution when the soft SUSY
breaking terms are taken at the GUT scale
\cite{[AM]}. Also this contribution does not involve any right handed down type
quark-squark mixing, so that it has the same phase structure as the SM does.
Consequently, the CKM measurement is not affected from the charged Higgs
contribution as we will discuss later. Chargino and neutralino contributions are
usually small
\cite{[AM],[CK]} and have no effect on the CKM measurements. 
 
The soft terms (e.g A and or
$\mu$) can also be complex. In that case one can get phases in $\Delta m_{B}$
even without grand unification. The complex terms in the mass matrix for the
squarks and sleptons are then responsible  for the  new phases which are
somewhat restricted by the edm of electron or neutron \cite{[HG]}, however large
phases can appear when the scalar masses are in the TeV range
\cite{[TF]}. There could also  be an induced phase in A due to the phase in the
Yukawa sector through renormalization, even when A is real at the GUT
scale. The phase induced is really small and gives rise to  the edm of electron
well within the experimental limit for squarks and gluino masses O(100 GeV)
\cite{[VB]}. It is possible to get comparable
$\Delta m_{B_d}$ and
$\Delta m_{K}$ from supersymmetric contribution with new phases \cite{[MW]} in a
model based on the MSSM (without grand unification) with right handed mixing
matrix in the up sector.  

The contribution to the $\Delta m_{B_{d,s}}$ can be parameterized as :
\begin{eqnarray}
\Delta m_{B_d}&=&A_{SM}+B_{SUSY}+C_{SUSY} e^{i\phi},\\\nonumber
\Delta m_{B_d}&=&A_{B_d} V_{td}^{*^2}V_{tb}^2 e^{i\phi_{B_d}}.
\end{eqnarray} To make the analysis a most general one we have included
$B_{SUSY}$ which has the same phase structure as A$_{SM}$. In our example, for
$\Delta m_{B_d}$, the box diagram with the LRLR structure (helicities of the
fermions in the external legs) has the mixing structure
$\left|V_{td}\right|^2 e^{i\phi}$,
 and RRRR type of box diagram has the mixing structure
$V_{td}^{^2} e^{2i\phi}$ in the diagonal quark mass basis with just b squark in
the loop, where
$\phi$ arises from the matrix $S$.  Note that even if $\phi$ is 0, both the RRRR
type and LRLR type still have different phase structures compared to the SM. As
a matter of fact any contribution from beyond the SM including multi-Higgs
models and left-right models can be written as above.
$A_{B_d}e^{i\phi_{B_d}}$ originates from the combination of the SM contribution
and the new contribution. Similarly we have for $B_s-{\bar B_s}$ and $K-{\bar
K}$ mixing:  
\begin{equation}
\Delta m_{B_s}=A_{B_d} V_{ts}^{*^2}V_{tb}^2 e^{i\phi_{B_s}}, \; \Delta
m_{K}=A_{K} V_{cs}^{*^2}V_{cd}^2 e^{i\phi_{K}}.
\end{equation}  

Expressions for ${q/ p}$ for each of these mesons are now:
\begin{equation} \left({q\over p}\right)_{B_d}=\left({V_{tb}^{*}{V_{td}}\over
{V_{tb}V_{td}^*}}\right)e^{-i\phi_{B_d}}, \; \left({q\over
p}\right)_{B_s}=\left({V_{tb}^{*}{V_{ts}}\over
{V_{tb}V_{ts}^*}}\right)e^{-i\phi_{B_s}}, \;
\left({q\over p}\right)_{K}=\left({V_{cd}^{*}{V_{cs}}\over
{V_{cd}V_{cs}^*}}\right)e^{-i\phi_{K}}.
\end{equation} In general $\phi_{B_d} $, $\phi_{B_s} $ and $\phi_K$ are
unrelated.  These phases are so defined that they are in addition to the phases
present in the SM, and can be treated as separate observables.  Charged Higgs
mediated box diagram has $C=0$, and CKM measurements are unaltered.   However,
in our calculation $C$ is non-zero (the LRLR type and the RRRR type), and will
affect  CKM phase measurements.

The CKM phases are defined as:
$\alpha = \mbox{Arg} \left( -{V^*_{tb} V_{td}/V_{ub}^* V_{ud}} \right),$
 $\beta = \mbox{Arg} \left( -{V^*_{cb} V_{cd}/V_{tb}^* V_{td}} \right),$ 
 $\gamma = \mbox{Arg} \left( -{V^*_{ub} V_{ud}/V_{cb}^* V_{cd}} \right)$. 
 Based on the SM, many methods have been suggested for measuring these CKM phases
using $B$ decays \cite{a,bb,d,e}. The cleanest method involves time-dependent 
measurements of rate asymmetries in neutral
$B$ decays to CP eigenstates \cite{a}, where one measures the time-dependent
rate asymmetry
$a_{f_{CP}}(t)$ which is a function of $\lambda \equiv \left({q/p}\right)_i
\left( {\bar A/A}
\right) $,    where $i$ denotes the corresponding mixing, i.e., $i= B_d, B_s$,
or $K$, and   
$A \equiv A(B^0 \rightarrow f)$ and $\bar A \equiv \bar A(\bar B^0 \rightarrow
f)$ with CP eigenstate $f$.

We shall analyze the different CP eigenstates that have been suggested, and
consider carefully what phases the measurements now yield. Our assumption for
decay amplitudes is that, while the tree amplitudes have the SM phases, any loop
process could have an additional unknown phase arising from beyond the Standard
Model.  Thus for penguin amplitudes we have 
\begin{eqnarray}
 {\bar A \over A} = \left({\bar A \over A}\right)_{SM} e^{i\phi_{peng}},
\end{eqnarray}  where $\phi_{peng}$ is a phase in addition to SM phase.   The
results of our analysis are presented in a convenient tabular form (Table I)
modeled after a similar table in the SM given in \cite{f}. Some of the modes
have also been discussed in the reference \cite{[BHS]} where only the SUSY
grandunification contributions are retained.  It is important to realize that
with our definitions of additional phases as defined in Eqs. (4) and (6), these
phases are measurable.  Further, the analysis is essentially model independent,
as these new phases can arise in any model beyond the SM. 

In row (1) we consider $B_d \rightarrow \psi K_S$.  This mode which is tree
dominated has
$Im\lambda$ given by: 
\begin{eqnarray} 
 Im \lambda &=& Im \left[ \left( {q \over p} \right)_{B_d} \left( {q \over p}
\right)_K 
                        \left( {\bar A \over A} \right) \right]
  = Im \left[ \left( {V^*_{tb} V_{td} \over V_{tb} V^*_{td}}
\right)e^{i\phi_{B_d}} 
              \left( {V_{cs} V^*_{cd} \over V^*_{cs} V_{cd}}
\right)e^{i\phi_{K}} 
              \left( {V_{cb} V^*_{cs} \over V_{cb}^* V_{cs}} \right) \right]
\nonumber \\
  &=& - \sin(2\beta + \phi_{B_d} + \phi_{K}). 
\label{beta} 
\end{eqnarray}    Note that the mode $b\rightarrow c{\bar c}s$ has negligible
penguin contribution.  In the SM this measurement yields $\sin (2\beta)$. 
Similarly $B_s \rightarrow \psi \phi$,
$\psi \eta$ would yield $\phi_{B_s}$ while in the SM there is no asymmetry.  In
row (2) and (4) we have pure penguin processes
$b\rightarrow s{\bar s}s$ and $b\rightarrow d{\bar d}s$, respectively.  These
could have an additional weak phase $\phi_{peng}$ or
$\phi_{peng}'$  corresponding to each process.  In row (2) the weak phases in
$B_s$ and
$B_d$ are the same $\phi_{peng}$ because they arise from the same quark
subprocess. The processes in row (3) are generally not suitable  as both tree
and penguin amplitudes make comparable contributions to the final states. In row
(5) tree amplitude dominates and although the modes are Cabibbo suppressed, they
are useful. In row (6) it is assumed that in the SM top contribution dominates
in the loop. The contributions from charm and up quarks are expected to be about
10\% over most of the allowed range \cite{c}. In row (7) tree contribution
dominates and the small penguin admixture can be removed using isospin analysis
\cite{bb}. Row (8) has processes dominated by tree diagrams and even though the 
mode
$D^0 K^*$ is not a CP eigenstate, an analysis of this mode can be used to
determine
$\gamma$ \cite{d}.  The charged $B$ decay mode $D^0 K^+$ can be used
alternatively, based on the same type of analysis \cite{e}.  

It is clear from the table I that from $B_d$ decays we can extract the
combination $\beta +\phi_{B_d}/2$ and $\phi_K$, $\phi_{peng}$ and $\gamma$. From
$B_s$ decays it is possible to measure
$\phi_{B_s}$, $\phi_K$, $\phi_{peng}$, $\phi_{peng}'$ and $\gamma$ and  the
combination $\beta +\phi_{peng}''/2$. However, combining both measurements, it
is possible in principle to extract all phases separately. Thus $\beta$ and
$\gamma $ are determined and $\alpha$ can be solved for. Since all the
measurements involve $sine$ of some angle, there exists some ambiguity in
determination of a definite angle.  However the analysis involved in the process
$D^0 K^*(892)$  is in principle expected to determine the definite value
$\gamma$, if in addition one  studies the exclusive processes $B_d \rightarrow
D^0 X^0$ ($X^0$ is $K^+ \pi^-$, $K^+
\pi^- \pi^0$, etc.) to remove discrete ambiguity \cite{d}. 

We recall that in the SM with three generations, the sum of three CKM phases
$\alpha$, $\beta$ and $\gamma$ must be equal to $\pi$.  In order to check the
validity of this  unique feature, one would measure the CKM phases, for
instance, through $B_d$ decay modes such as
$\pi \pi$, $\psi K_S$ and $D^0 K^*$(892) which are preferred experimentally and
would yield $\alpha$, $\beta$  and $\gamma$, respectively, in the SM.   However,
as we can see from Table I, these modes would actually measure
$\pi-(\beta +\gamma +\phi_{B_d}/2)$, $\beta +(\phi_{B_d} +\phi_K)/2$ and
$\gamma$, respectively.   The sum of these three angles would give
$\pi+\phi_K/2$ which can be a good indication for new  physics unless $\phi_K$
turns out to be small.  Even in case the experiments show the sum of these
angles to be $\pi$, there is still room left for extra physics because of the 
possible existence of
$\phi_{B_d}$ or $\phi_{B_s}$. Another interesting case is of multi-Higgs models,
where SM phases might be absent. This corresponds to $\gamma=\beta=0$,
$\alpha=\pi$. In that case, asymmetry in
$B_d\rightarrow \psi K_s$ is opposite in sign to $B_d\rightarrow \pi\pi$, and
$\gamma$ measurement will yield 0.

If we concentrate just on the $B_d$ decay modes, since these decay modes are
more preferable from the experimental viewpoint, it is hard to extract all the
CKM angles cleanly. But the angle $\gamma$  can still be measured without
contamination of the extra phases.  Since it seems to be very  difficult to
extract $\alpha$ and $\beta$ by using any $independent$ methods, we suggest that
$\alpha$ and $\beta$ be determined using the unitarity triangle.  Measuring the
ratio of the CKM factors $|V_{ud} V_{ub}| / |V_{cd} V_{cb}|$ (e.g. by studying
the spectra of charged leptons in the semileptonic processes $b \rightarrow u
\bar
\nu_e e$ and $b \rightarrow c \bar \nu_e e$) and using 
$\gamma$ when measured, one can construct the unitarity triangle completely,
which enables one to determine the phases 
$\alpha$ and $\beta$ simultaneously. This angle $\beta$ should be compared with
the angle measured in the $B_d$ decay modes such as $\psi K_S$ in order to
extract information about new physics. 

In conclusion, we have shown how the measurement of CKM phases as well as
additional phases can be achieved when comparable contributions from beyond the
standard model might be present.

This work was supported by Department of Energy grant DE-FG03-96ER 40969.  
 
\newpage
\begin{table}
\caption{$B$ decay modes for measuring CP angles.}
\begin{tabular}{cccccc}  
  & Quark Process  & $B_d$ Modes  & $B_d$ Angles  & $B_s$ Modes  & $B_s$ Angles 
\\ \hline  (1) & $b
\rightarrow c \bar c s$ & $\psi K_S$ & $\beta +(\phi_{B_d} +\phi_K)/2$ & 
 $\psi \eta$, $\psi \phi$, & $\phi_{B_s}/2$    \\ 
    &  &  &  & $D^+_s D^-_s$ &     \\  \hline   (2) & $b \rightarrow s \bar s s$
& $\phi K_S$ &
$\beta +(\phi_{B_d} +\phi_K +\phi_{peng})/2$ & 
 $\phi \eta$ & $(\phi_{B_s} +\phi_{peng})/2$    \\   \hline  (3) & $b
\rightarrow u \bar u s$ &
$\pi^0 K_S$, $\rho^0 K_S$ & $-$ & 
 $\phi \pi^0$, $K^+ K^-$ & $-$    \\  \hline  (4) & $b \rightarrow d \bar d s$ &
$\pi^0 K_S$,
$\rho^0 K_S$ & $-$ & 
 $K^0 \bar K^0$ & $(\phi_{B_s} +\phi'_{peng})/2$   \\  \hline  (5) & $b
\rightarrow c \bar c d$ &
$D^+ D^-$, $\psi \pi^0$, & 
 $\beta +\phi_{B_d}/2$ & $\psi K_S$ & $(\phi_{B_s} +\phi_K)/2$    \\ 
    &  & $D^0 \bar D^0$  &  &  &     \\  \hline  (6) & $b \rightarrow s \bar s
d$ & $K^0 \bar K^0$ &
$(\phi_{B_d} +\phi''_{peng})/2$ & 
 $\phi K_S$ & $\beta +(\phi_K + \phi_{B_s} + \phi''_{peng})/2$    \\  \hline  (7)
&
$b
\rightarrow u \bar u d$, &
$\pi \pi$, $\pi \rho$, & $\pi -(\beta +\gamma +\phi_{B_d}/2)$ & 
 $\rho^0 K_S$, $\pi^0 K_S$ & $\gamma +(\phi_{B_s} +\phi_K)/2$    \\
    & $d \bar d d$ & $\pi a_1$ &  &  &    \\   \hline  (8) & $b \rightarrow c
\bar u s$, & $D^0_{CP} K^*$(892) & $\gamma$ & 
 $D^0_{CP} \phi$ & $-$    \\
    & $u \bar c s$ & $(D^0_{CP} K^+)$ &  &  &    \\  
\end{tabular}
\end{table}

\begin{figure}[htb]
\vspace{1 cm}

\centerline{ \DESepsf(susy1.epsf width 12 cm) }
\smallskip
\caption {Plots of r($\equiv\Delta m_{{B}_{gluino}}$/$\Delta m_{{B}_{SM}})$ as a
function of $\mu $ for different values of the gaugino masses ($m_{1/2}$). The
scalar mass $m_0$=1 TeV for the entire plot.}
\vspace{1 cm}

\centerline{ \DESepsf(susy22.epsf width 12 cm) }
\smallskip
\caption {Plots of r($\equiv\Delta m_{{B}_{gluino}}$/$\Delta m_{{B}_{SM}}$) as a
function of the gaugino mass ($m_{1/2}$) for different values of the scalar
masses $m_0$.}
\vspace{3 cm}

\end{figure}

\end{document}